\documentclass[%
 reprint,
preprintnumbers,
nofootinbib,
 amsmath,amssymb,
 aps,
]{revtex4-2}

\usepackage{graphicx}
\usepackage{dcolumn}
\usepackage{bm}
\usepackage[colorlinks=true,linkcolor=blue,citecolor=blue,urlcolor=blue]{hyperref}
\usepackage[mathlines]{lineno}
\usepackage{here}

\usepackage{color}
\newcommand{\ikeda}[1]{\textcolor{black}{#1}}

\begin{document}

\preprint{APS/123-QED}

\title{Background contributions in the electron-tracking Compton camera onboard SMILE-2+}

\author{Tomonori Ikeda}
\email{ikeda.tomonori.62h@st.kyoto-u.ac.jp}
\author{Atsushi Takada}
\author{Taito Takemura}
\author{Kei Yoshikawa}
\author{Yuta nakamura}
\author{Ken Onozaka}
\author{Mitsuru Abe}
\author{Toru Tanimori}
\affiliation{Graduate School of Science, Kyoto University Kitashirakawa Oiwakecho, Sakyo, Kyoto, Kyoto, 606-8502, Japan}

\author{Yoshitaka Mizumura}
\affiliation{Institute of Space and Astronautical Science, Japan Aerospace Exploration Agency Yoshinodai 3-1-1, Chuou, Sagamihara, Kanagawa, 252-5210, Japan}






\date{\today}

\begin{abstract}
Mega electron volt (MeV) gamma-ray observations are promising diagnostic tools for observing the universe. 
However, the sensitivity of MeV gamma-ray telescopes is limited by peculiar backgrounds, restricting the applicability of MeV gamma-ray observations. Thus, background identification is crucial in the design of next-generation telescopes. 
Here, we assessed the background contributions of the electron-tracking Compton camera (ETCC) onboard SMILE-2+ in balloon experiments.
This assessment was performed using Monte Carlo simulations.
The results revealed that a background below 400~keV existed due to the atmospheric gamma-ray background, cosmic-ray/secondary-particle background, and accidental background.
Moreover, an unresolved background component that was not related to direct Compton-scattering events in the ETCC was confirmed above 400~keV.
Overall, this study demonstrated that the Compton-kinematics test is a powerful tool for removing backgrounds and principally improves the signal-to-noise ratio at 400~keV by an order of magnitude.

\end{abstract}

\maketitle


\section{Introduction} \label{sec:intro}
Mega electron volt (MeV) gamma-ray observations have been used to investigate various unresolved issues in modern astrophysics, such as the particle acceleration processes of relativistic jet and outflow sources~\cite{Chiaberge2001}, the origin and propagation of low-energy cosmic rays associated with star formation~\cite{Grenier2015}, and the nucleosynthesis and chemical enrichment of our galaxy~\cite{Diehl2013}.
In particular, the morphology of a bright gamma-ray $e^{+}e^{-}$ annihilation line cannot be easily explained using conventional astrophysical sources, such as type ${\rm I}$a supernovae, massive stars, microquasars, and X-ray binaries, and has thus remained an issue in sub-MeV observations~\cite{Prantzos2011,Siegert2016}.
Recently, Advanced LIGO and Virgo established the foundation of gravitational wave astronomy~\cite{LIGO2016}. 
In addition, IceCube~\cite{Aartsen_2016} has detected astrophysical neutrinos. 
The astrophysical sources of gravitational waves and high-energy neutrinos are expected to emit high-energy gamma rays~\cite{Takami2014,IceCube2016}.
Therefore, coincidence observations of gamma-ray signals in time and space are desired to obtain complete and complementary information for new astronomy concepts in the multimessenger epoch.

The Compton telescope COMPTEL onboard the Compton gamma-ray observatory has been the most successful experiment in the field of MeV gamma-ray observations~\cite{1993COMPTEL}. 
The novel tools employed for background reduction in COMPTEL can measure the time-of-flight (ToF) and the pulse shape discriminator (PSD).
The ToF information aids in distinguishing forward-scattered and backward-scattered events, and the PSD enables the rejection of neutron scattering events in organic scintillators~\cite{Weidenspointer2001}.
\ikeda{Although the COMPTEL telescope is an ingenious device, it has only moderate sensitivity due to unexpectedly high background contamination and thus requires the application of strict data cuts such as the so-called Earth horizon cut.}
In practice, the achieved sensitivity requires an observation period that is 4.5 times longer than the predicted duration.
\ikeda{Arguably, the next generation of Compton telescopes should include other comparable background-rejection features, such as consistency checks for the Compton kinematics measuring either the direction of the Compton recoil electron~\cite{2004Schonfelder, Tanimori2004} or multiple Compton interactions~\cite{BEECHERT2022166510}.} 

Therefore, we proposed the sub-MeV/MeV gamma-ray imaging loaded-on-balloon experiment (SMILE)~\cite{Tanimori2004, Tanimori2015}) using an electron-tracking Compton camera (ETCC), which can record all information regarding Compton kinematics, including the direction of the Compton electron. 
The ETCC enables gamma-ray imaging based on geometrical optics~\cite{Tanimori2017,Bernard2022}.
A second balloon experiment, hereafter referred to as SMILE-2+, was conducted to observe the Crab Nebula in 2018. 
We successfully obtained a significance of 4.0${\sigma}$ in the energy range of 0.15--2.1~MeV~\cite{2022Takada} and acquired data for approximately one day at high altitudes, which were utilized for background validation.

\begin{figure}[h]
\includegraphics[width=8.6cm]{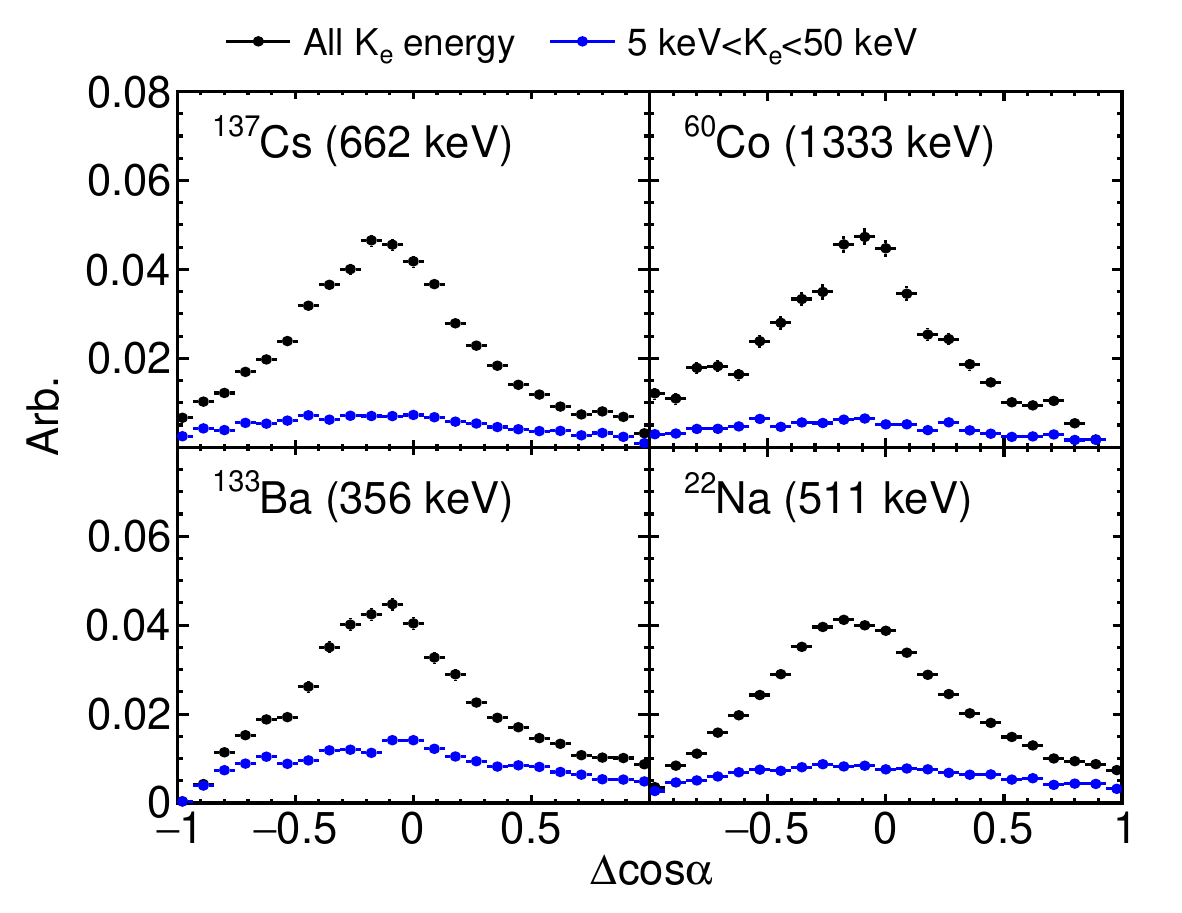}
\caption{
\ikeda{Calibration data of $\Delta\cos\alpha$ obtained using $^{137}$Cs (top-left panel), $^{60}$Co (top-right panel), $^{133}$Ba (bottom-left panel), and $^{22}$Na (bottom-right panel). 
The black points show unselected recoil electron events. The blue points represent selected events with 5~keV$<K_{ {\rm e} }<$50~keV.}
}
\label{fig:alpha_calib}
\end{figure}

The ETCC includes a scattering and absorption medium similar to the medium used in the conventional Compton camera.
We expect that the gamma-ray initially loses the partial energy in the scattering medium via Compton scattering interactions, followed by complete absorption in the absorption medium, hereafter referred to as the ideal Compton event. 
However, several undesired events can occur, such as the double Compton scattering event, the back-scattering event in which the gamma-ray was initially scattered on the absorption medium, and the accidental event in which two photons interacted in each medium.
The ETCC provides a powerful tool to discriminate ideal Compton events from undesired events: a consistency check of the Compton kinematics based on the direction information of the recoil electron.
We can calculate the differential angle $\alpha$ between the scattering gamma-ray $\mathbf{g}$
\footnotemark[1] \footnotetext[1]{$\mathbf{e}$ and $\mathbf{g}$ are unit vectors.} and the recoil electron $\mathbf{e}$ with two approaches. The first is a geometrical calculation, which is formulated as
\begin{equation}
    \cos\alpha_{{\rm geo}} = \mathbf{g}\cdot \mathbf{e}.
\end{equation}
The second is a Compton kinematics calculation, which is formulated as follows:
\begin{equation}
    \cos\alpha_{{\rm kin}} = \left(1-\frac{m_{{\rm e}}c^{2}}{E_{\gamma}}\right) \sqrt{\frac{K_{{\rm e}}}{K_{{\rm e}}+2m_{{\rm e}}c^{2}}},
\end{equation}
where $m_{{\rm e}}$ denotes the electron mass, $E_{\gamma}$ represents the energy of the scattered gamma-ray and $K_{{\rm e}}$ represents the kinetic energy of the recoil electron.
The difference value $\Delta\cos\alpha~(=\cos\alpha_{{\rm geo}}-\cos\alpha_{{\rm kin}})$ is crucial for identifying ideal Compton events which yield $\Delta\cos\alpha=0$.
In practical scenarios, because the measurement of such parameters involves uncertainties associated with the resolution, the $\Delta\cos\alpha$ distribution of the Compton event is broadened. 
\ikeda{
Fig.~\ref{fig:alpha_calib} shows the calibration data of $\Delta\cos\alpha$ obtained using various gamma-ray sources. Here, the black and blue points represent the unselected electron energy data and the selected data of 5~keV$<K_{{\rm e}}<$50~keV, respectively.
In the ETCC onboard SMILE-2+, poor determination accuracy for the recoil direction of the low energy electron leads to the broadening of the $\Delta\cos\alpha$ distribution~\cite{Ikeda2021}.}
However, the shape of the $\Delta\cos\alpha$ distribution provides sufficient information to ensure the detection of the Compton scattering event.

In this study, we reanalyzed the SMILE-2+ data and elucidated the background contribution using the Monte Carlo simulation.
Furthermore, we confirmed that the additional parameter $\Delta\cos\alpha$ can aid in identifying ideal Compton events.
The remainder of this paper is divided into four sections.
In Section~\ref{sec:smile}, the SMILE-2+ balloon flight experiment is introduced, and the background dataset used to evaluate the simulation results is established.
In Section~\ref{sec:bg_simulation}, the SMILE instrument and the mass model used in the Monte Carlo simulations are briefly described. 
Furthermore, the radiation environment at the balloon altitude and the general characteristics of the background component are specified. 
In Section~\ref{sec:result}, the experimental data and background simulation results are compared, and the background contribution to SMILE-2+ is discussed.
Finally, the Compton kinematics test results are validated in Section~\ref{sec:discussion}.

\section{SMILE-2+ balloon flight and dataset}\label{sec:smile}
\begin{figure}[h]
\includegraphics[width=8.6cm]{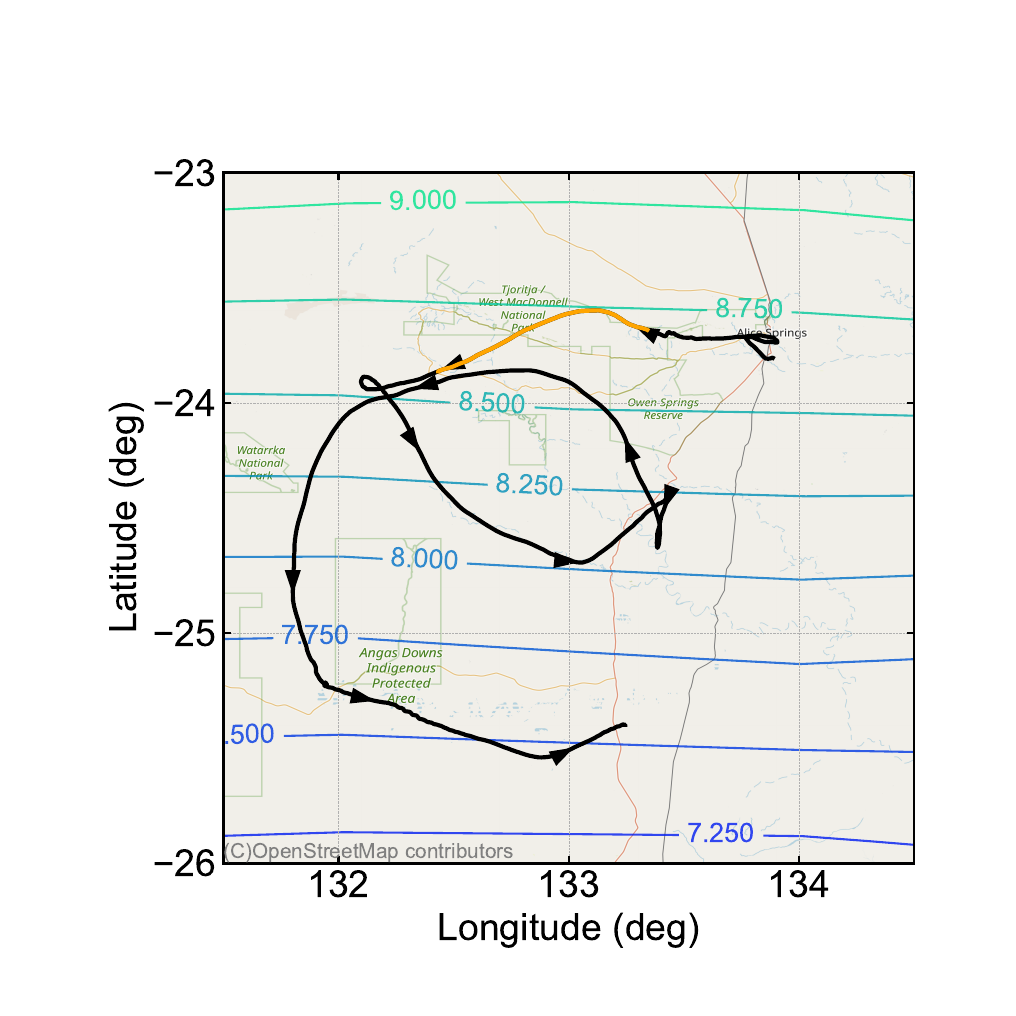}
\caption{Flight trajectory of the SMILE-2+ balloon launched from Alice Springs at 06:24 ACST on April 7, 2018, until the end of the operation at 10:53 ACST on April 8, 2018. 
The orange line shows the selected data, which were used as steady background observations. The contour lines show the cutoff rigidity calculated via PARMA.}
\label{fig:flight_path}
\end{figure}

The flight trajectory of the SMILE-2+ balloon is shown in Fig~\ref{fig:flight_path}.
The balloon was successfully launched from the Australian balloon launch station, Alice Springs, on April 7, 2018, at 06:24 Australian Central Standard Time (ACST). 
The balloon attained a floating altitude of 39.6~km after an ascent time of 2~h and the operation was terminated at 10:53 ACST on April 8, 2018. The total duration of the high-altitude observation was approximately 26~h.

\begin{figure}[h]
\includegraphics[width=8.6cm]{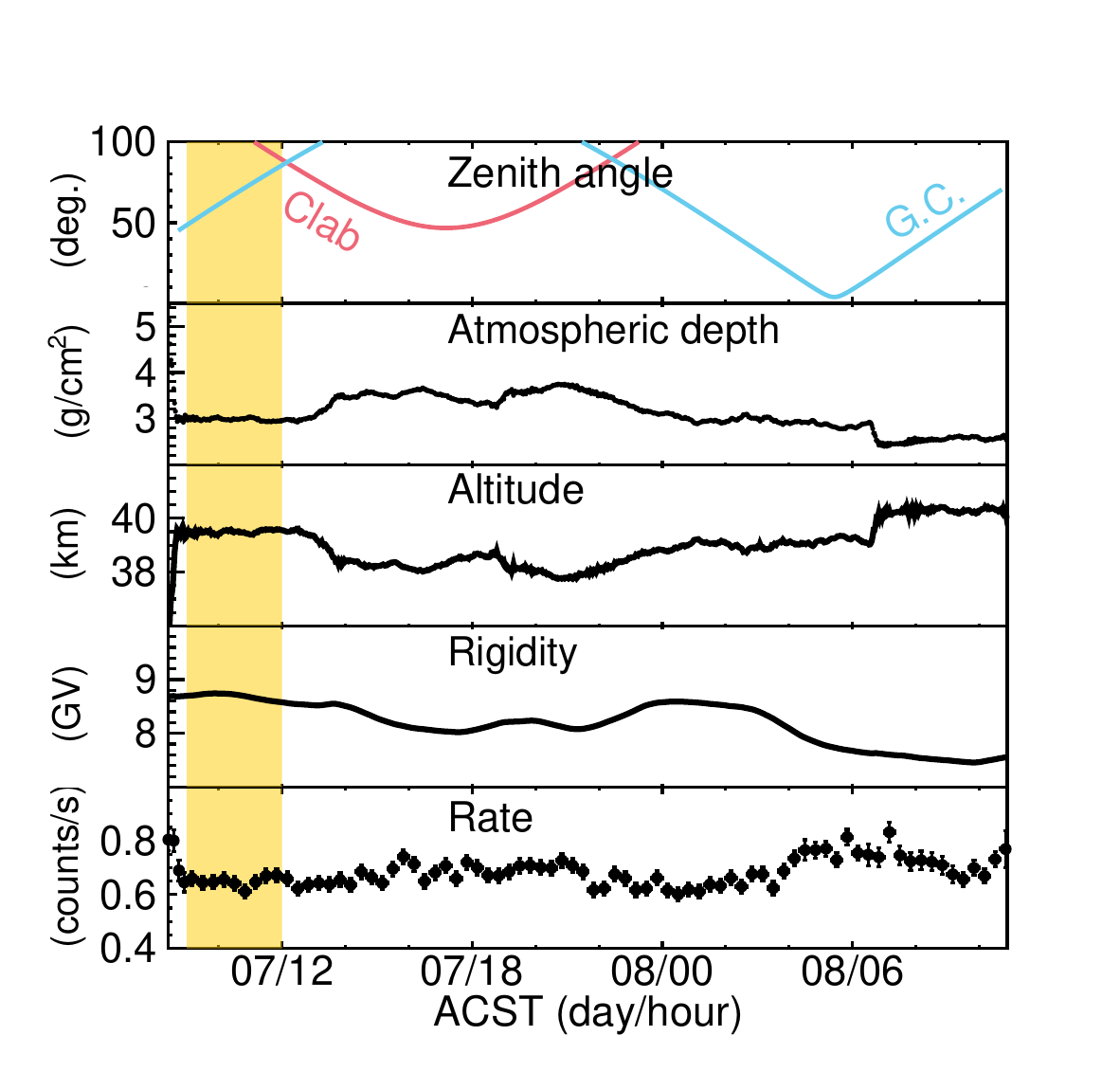}
\caption{The first panel shows the zenith angle of the bright sources.
The second panel shows the time variation in the atmospheric depth. 
The third and fourth panels show the time variation of the altitude and vertical cutoff rigidity calculated via PARMA, respectively. 
The last panel presents the measured count rate with the gamma-ray selection in the FoV.
The orange-shaded region indicates the dataset selected as the steady background observation.}
\label{fig:condition}
\end{figure}

The atmospheric depth and the vertical cutoff rigidity varied slightly as the balloon floated at high altitudes. Fig.~\ref{fig:condition} shows the time variations of the zenith angle of the bright sources, the atmospheric depth, the vertical cutoff rigidity calculated by PARMA~\cite{PARMA}, and the measured count rate in the energy range of 150--2100~keV. 
The decrease in altitude (increase in atmospheric depth) from 12:00 to 08:00 ACST can be attributed to sunset. 
Fluctuations in the atmospheric depth and the cutoff rigidity can lead to alterations in the number of atmospheric gamma rays, cosmic rays, and secondary particles induced by interactions between cosmic rays and atmospheric materials.
In reality, the measurement rate which was adopted by the gamma-ray selection~\cite{2022Takada} except for the Compton kinematics test in the field of view (FoV), corresponding to a zenith angle below 60$^{\circ}$, includes the atmospheric gamma-rays and the background gamma rays induced by the cosmic rays and the secondary particles. 
This parameter was correlated to the air mass and cutoff rigidity. 
Therefore, the dataset obtained from 09:00 to 12:00 on April 7, 2018, was defined as the steady background data and analyzed in the background validation.
This duration included no bright sources such as the Crab Nebula or the galactic center in the FoV.
The time-averaged altitude, atmospheric depth, vertical cutoff rigidity, and count rate of the steady background data were 39.5~$\pm$~0.06~km, 2.98~$\pm$~0.06~g/cm$^{2}$, 8.69~$\pm$~0.05~GV, and 0.65~$\pm$~0.01~count/sec, respectively.

\section{Background simulations}\label{sec:bg_simulation}
Background simulations were performed using Monte Carlo simulations based on Geant4~\cite{Geant4}. 
The simulation tool was optimized for SMILE-2+. 
The detector response of the ETCC, such as the effective area and the point spread function, reproduced the experimental data~\cite{2022Takada}.
We implemented a deep-learning method to improve the reconstruction accuracy of the recoil direction of the electron and the scattering position~\cite{Ikeda2021}.
The point spread function was improved by 32\% compared with the conventional method~\cite{2022Takada}.
In addition, we adopted the ANNRI-Gd model to simulate the gamma-ray energy spectra of the thermal neutrons captured on $^{157}$Gd and $^{155}$Gd~\cite{ANNRI2019,ANNRI2020}.
In this section, we describe the mass model used in the simulations and the environmental radiation related to the simulated particles. 
Finally, the types of background events triggering the ETCC detector are categorized for convenience.

\subsection{Instrument and mass model}
The SMILE-2+ instrument, including the detector performance, is described in Ref.~\cite{2022Takada}. 
Briefly, the ETCC detector onboard SMILE-2+ includes a time projection chamber (TPC) functioning as a Compton-scattering target and pixel scintillator arrays (PSAs) acting as gamma-ray absorbers. 
The TPC is filled with an argon-based gas (95\%Ar~$+$~3\%CF$_{4}$~$+$~2\%iso-C$_{4}$H$_{10}$). 
A micropattern gas detector, $\mu$-PIC~\cite{uPIC} with 768~$\times$~768 strips, is mounted on top of the TPC. 
The detection volume of the TPC is 30~$\times$~30~$\times$~30~cm$^{3}$. 
The TPC is surrounded by 108 PSAs. 
Each PSA is made of Gd$_{2}$SiO$_{5}$:Ce (GSO) scintillators of 8~$\times$~8 pixels with a pixel size of 6~$\times$~6 mm$^{2}$. 
The thicknesses of the bottom and side PSAs are 26~mm and 13~mm, respectively.

\begin{figure}[!h]
\includegraphics[width=8.6cm]{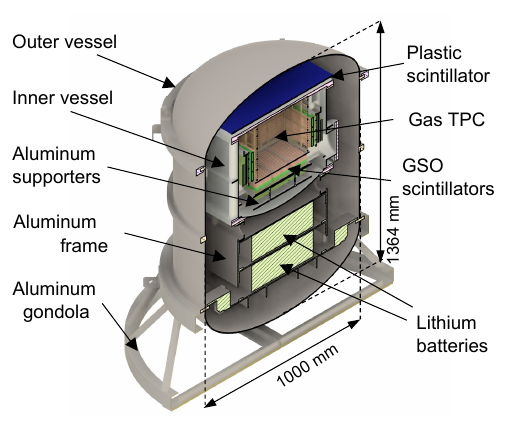}
\caption{SMILE-2+ mass model. 
}
\label{fig:mass_model}
\end{figure}

We developed a detailed mass model corresponding to the SMILE-2+ instrument to obtain a reliable detector response in the Monte Carlo simulations. Fig.~\ref{fig:mass_model} shows a schematic of the constructed SMILE-2+ mass model. 
An aluminum outer vessel was located on an aluminum gondola. 
To remove charged cosmic rays, a veto system consisting of a 5~mm--thick plastic scintillator was used and placed on top of the TPC. 
The interior vessel, which was composed of stainless steel and aluminum, was used to cover the gas TPC and GSO scintillators. 
The GSO scintillators were supported by aluminum and Teflon frames. 
Two lithium batteries were installed under the inner vessel. 
A ballast of iron sand was placed at the bottom of the gondola (not included in Fig.~\ref{fig:mass_model}) to maintain the floating altitude. 
Table~\ref{tab:mass_model} summarizes the primary materials and masses in the instrument.
The total mass of the constructed model was 729.2~kg, which differed from the measured mass by 9.1\%.

\begin{table}[h]
\caption{Instrument materials in the mass model}
\label{tab:mass_model}
\begin{ruledtabular}
\begin{tabular}{lcc}
\textrm{Component}&
\textrm{Material}&
\textrm{Mass (kg)}\\ 
\colrule
Outer vessel & Al & 65.24\\
Aluminum gondola      & Al, Mg, and Si & 53.33\\
Inner vessel & Cr, Ni, and Al & 90.84\\
Aluminum supporters & Al & 10.77\\
Aluminum frame & Al, and Mg & 59.20\\
Lithium batteries & Li$_{10}$Ni$_{4}$Mn$_{3}$Co$_{3}$O$_{20}$ & 79.20 \\
Ballast & Fe & 325.98\\
Plastic scintillator & C$_{9}$H$_{10}$ & 2.14 \\
Gas TPC & Ar, CF$_{4}$, and iso-C$_{4}$H$_{10}$ & 0.10\\
GSO scintillators & Gd$_{2}$SiO$_{5}$, and Ce & 28.94\\
\end{tabular}
\end{ruledtabular}
\end{table}

\subsection{Radiation environment}\label{sec:radiation_environment}
Because the atmospheric opaque to gamma-ray photons obstructs the observation of comic MeV gamma rays, the detectors must be positioned at the top of the Earth's atmosphere by the balloon or into space. 
The instrument is exposed to variable radiation in such a high-altitude environment.
\ikeda{The main constituents of radiation include cosmic rays, secondary particles from the Earth's atmosphere, atmospheric gamma rays, internal radiation from the primordial radioactivity of the detector materials, and radioactive decay due to activation from hadron interactions.}
Furthermore, the radiation environment is concisely described below, followed by the background categorization.

\subsubsection{Cosmic rays and secondary particles}
Protons and helium nuclei are the most abundant components of cosmic rays, followed by electrons. 
Although such primary cosmic rays come from the exterior of the solar system, the solar wind magnetic field, commonly referred to as solar modulation, decelerates the incoming low-energy charged particles. 
\ikeda{
In addition, the Earth’s magnetic field prevents lower-energy charged cosmic-ray particles from reaching the balloon altitude.
}
Therefore, the intensity of the primary cosmic rays depends on solar activity and the cutoff rigidity.
Moreover, the primary cosmic rays can produce secondary elementary particles, such as protons, neutrons, electrons, positrons, and photons, through atmospheric interactions.
Because the intensity of the secondary particles is affected by the intensity of the primary particles and the atmospheric density, the computation time is inevitably large.

\begin{figure}[!h]
\includegraphics[width=8.6cm]{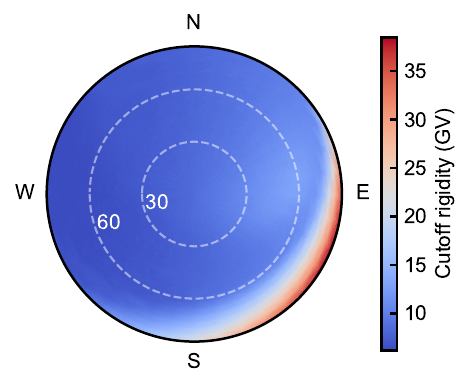}
\caption{Direction-dependent cutoff rigidity in the SMILE-2+ environment. West is toward the left, and South is toward the bottom of the image.}
\label{fig:rigidity_map}
\end{figure}

\begin{figure}[!h]
\includegraphics[width=8.6cm]{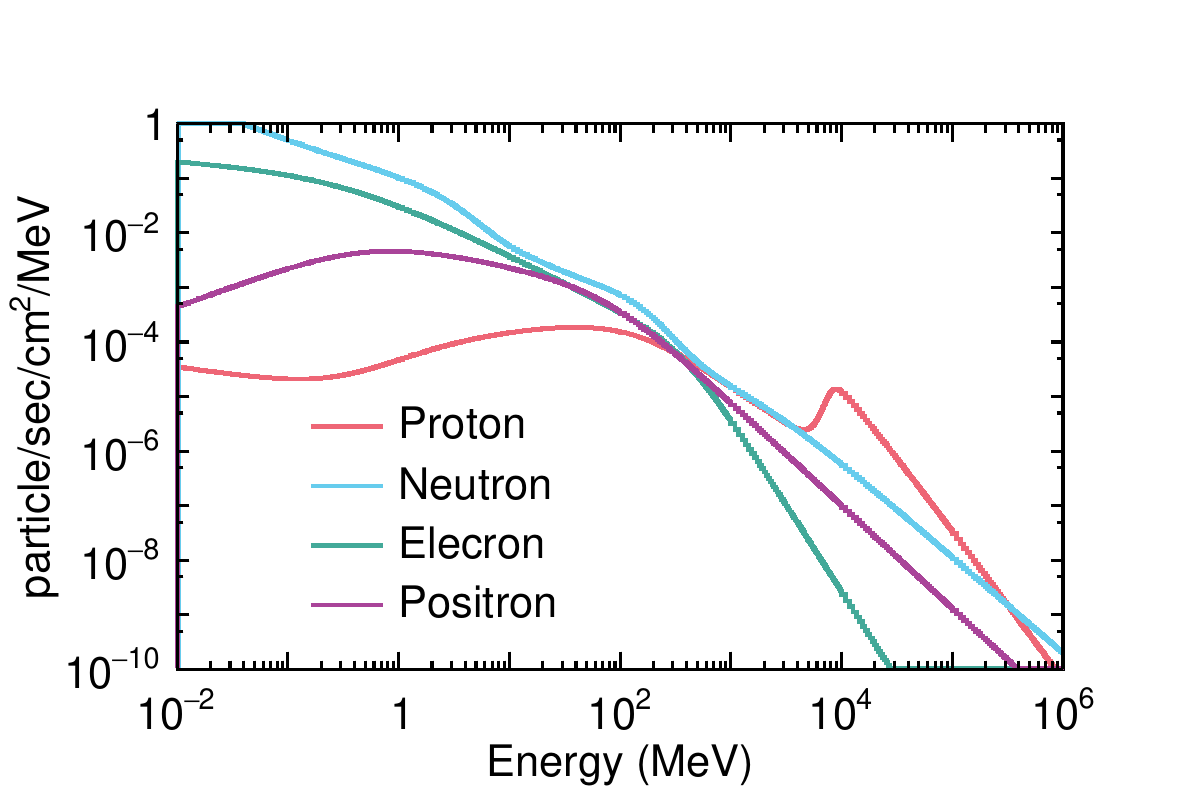}
\caption{Particle fluxes calculated using PARMA with a cutoff rigidity of 8.69~GV, an atmospheric depth of 2.98~g/cm$^{2}$, and the Wolf number of 16.5. 
The red, cyan, green, and purple lines indicate protons, neutrons, electrons, and positrons, respectively.}
\label{fig:flux_parma}
\end{figure}

Therefore, the intensity and the incident angular distribution of the cosmic rays and secondary particles were estimated using PARMA4.0 software~\cite{PARMA}. 
\ikeda{While there are simulation tools based on the Monte Carlo method like CORSIKA~\cite{CORSIKA}, COSMOS~\cite{Sako2021}, and FLUKA~\cite{FLUKA},
the PHITS-based analytical radiation model in the atmosphere (PARMA) was constructed using numerous analytical functions whose parameters were fitted to reproduce the results of the extensive air shower simulations performed with PHITS~\cite{PHITS}.}
Therefore, PARMA can instantaneously provide terrestrial cosmic-ray fluxes at various locations, altitudes, and solar activities. 
The direction-dependent cutoff rigidity in the SMILE-2+ environment, shown in Fig.~\ref{fig:rigidity_map}, was obtained using the COR tool~\cite{COR2022} and is roughly asymmetric with respect to the azimuth angle.
The vertical cutoff rigidity of 8.7~GV was consistent with that calculated using PARMA.
The azimuth and zenith dependence of the cutoff rigidity is considered in PARMA by assuming that the Earth's magnetic field is a simple dipole.
The calculated flux of protons, the secondary particles of the neutrons, electrons, and positrons in the SMILE-2+ background observation, is shown in Fig.~\ref{fig:flux_parma}.
The primary cosmic-ray protons lead to a peak at approximately 10~GeV in the proton spectrum.

\subsubsection{Atmospheric gamma rays}
Atmospheric gamma rays are produced in the Earth's atmosphere by cosmic rays. 
Low-energy continuum radiation below 10~MeV has been studied by Ling~\cite{Ling} and Sch\"{o}enfelder~\cite{Schoenfelder1977}.
High-energy radiation above 30~MeV has been investigated by Stecker~\cite{Stecker1973} and Thompson~\cite{Thompson}. 
Theoretically, the photon spectrum beyond 50~MeV is  predominantly produced by the decay of neutral pions induced in the nuclear collisions of cosmic rays. 
Below 50~MeV, the dominant process is the bremsstrahlung radiation of the relativistic electrons produced by charged pions and pair production~\cite{Beuermann}. 
The intensity model of the atmospheric gamma rays was developed by Ling~\cite{Ling}, Costa~\cite{Costa}, Morris~\cite{Morris}, and Sazanov~\cite{Sazanov}. 
Ling's model has been used in several balloon experiments and is compatible with experimental data~\cite{Schoenfelder1977,NCT2007,Takada_2011}. 
Thus, we employed Ling's model to estimate the atmospheric gamma-ray flux. 
However, notably, the upward gamma-ray flux of Ling's model differs from the measurements owing to Ling’s assumption of an isotropic gamma-ray source function in the atmosphere~\cite{Schoenfelder1977}. Furthermore, the energy range of the model is below 10~MeV. 
Therefore we used both Ling's model and the PARMA model.
The PARMA model is  derived from air shower simulations, while Ling's model was developed according to a semiempirical method based on measured gamma-ray fluxes.
This difference between the models leads to uncertainty in the detected spectrum and is discussed in Section~\ref{sec:result}.

\begin{figure}[!h]
\includegraphics[width=8.6cm]{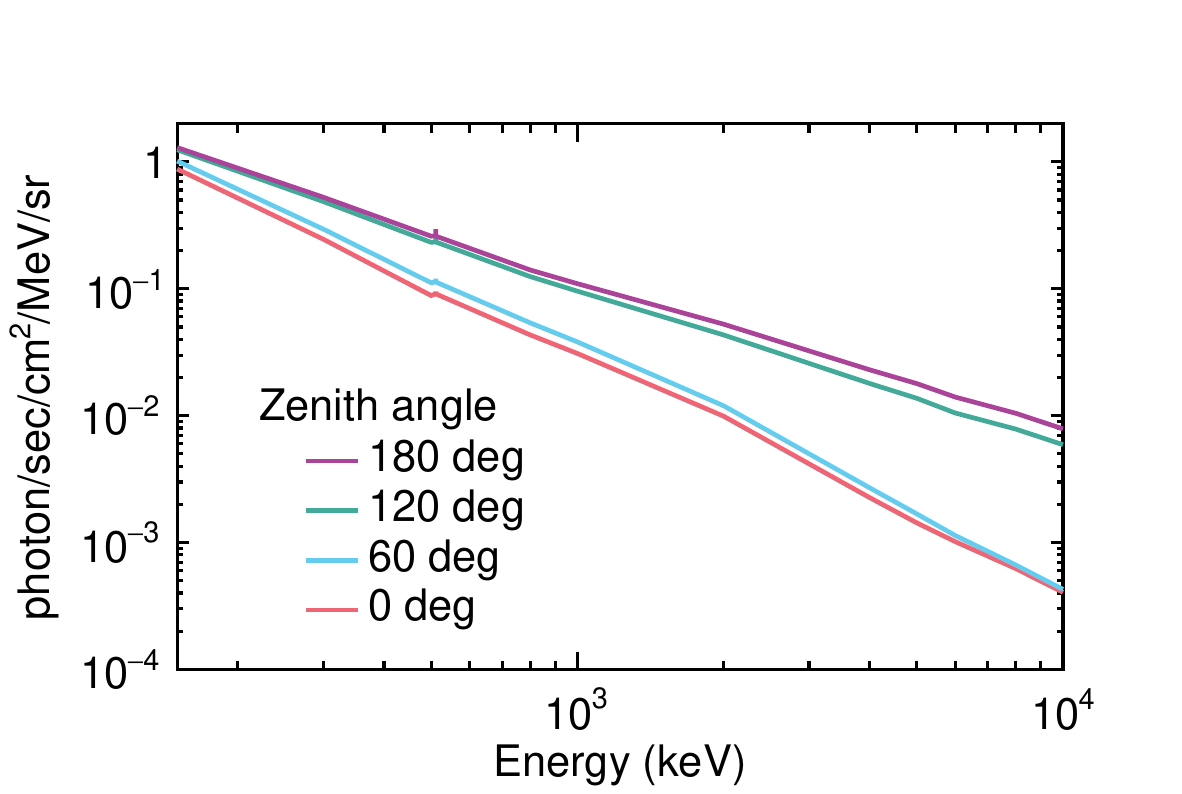}
\caption{Intensity of atmospheric gamma-rays calculated by the Ling model.
The purple, green, cyan, and red lines correspond to the intensity of the zenith angle at 180$^{\circ}$, 120$^{\circ}$, 60$^{\circ}$, and 0$^{\circ}$, respectively.}
\label{fig:flux_ling}
\end{figure}

We extracted the atmospheric gamma-ray component from Ling's model. 
Although the intensity of the atmospheric gamma rays depends on the cutoff rigidity and solar modulation, Ling's model is independent of these values. 
Considering this discrepancy between our environment and Ling's environment, we estimated the correction factor based on Sazanov's model, obtaining a correction factor of 0.89. 
The calculated atmospheric intensity in the SMILE-2+ environment is shown in Fig.~\ref{fig:flux_ling}.

\subsubsection{Internal radiation}\label{sec:internal_radiation}
It is well-known that many materials contain natural radioactive isotopes such as $^{238}$U and $^{232}$Th (U/Th). 
In particular, if materials near or in the detector contain considerable radioactive isotopes, the alpha, beta, and gamma rays in the decay series emitted by these isotopes can create the PSA triggers.
We used a high-purity germanium detector and identified that certain amounts of U and Th radioactive isotopes contaminated the GSO crystal in the PSA detector. 
The radioactivities of the U upper series and $^{176}$Lu corresponding to 3.4~$\pm$~0.4~Bq/kg and 89~$\pm$~5~mBq/kg, respectively, were detected. 
Because the U middle series, U lower series, and Th were not observed significantly, the 90\%-confidence level upper limits were obtained as 15~mBq/kg, 8.0~Bq/kg, and 3.7~mBq/kg, respectively.

\begin{figure}[!h]
\includegraphics[width=8.6cm]{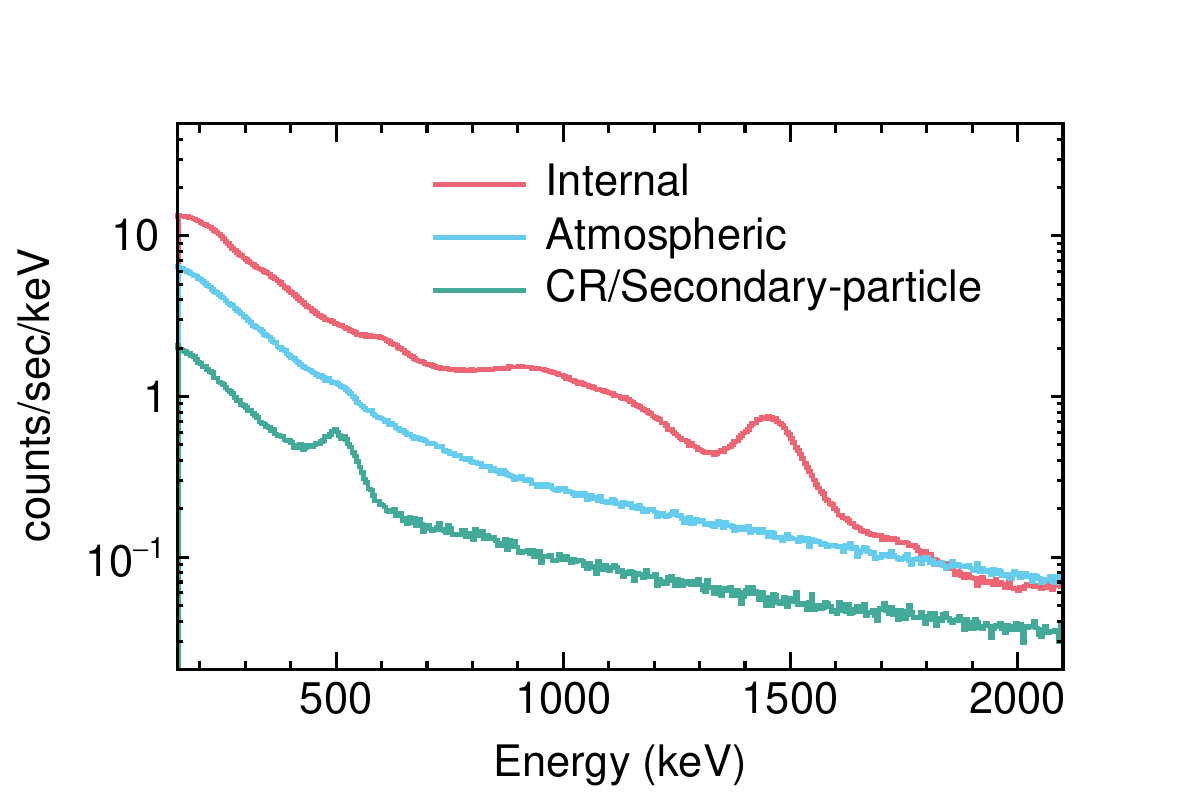}
\caption{Spectra of the PSA scintillator. 
The red line shows the internal radiation obtained in the ground experiment. 
The cyan and green lines denote the contributions from the atmospheric gamma rays and cosmic rays, respectively, including secondary particles and induced gamma rays.}
\label{fig:flux_internal}
\end{figure}

To investigate the characteristics of the energy spectrum in scenarios with such radioactive contamination, we acquired the data on the ground using the self-triggering mode of a PSA detector with lead shielding.
The obtained spectrum is shown in Fig.~\ref{fig:flux_internal}. 
The bump component at approximately 1~MeV was caused by the alpha emission from the decay chain of U/Th in the GSO crystal~\cite{WANG2002498,KAMAE2002456}. 
Moreover, we detected a 1.4~MeV line of $^{40}$K radioactivity.
The glass of the photomultiplier tubes (Hamamatsu Photonics, flat-panel H8500) was suspected as the potassium source. 
The total $^{40}$K radioactivity of the photomultiplier tubes was determined 10.4~$\pm$~0.1~Bq.

We simulated the expected spectra of the PSA scintillator from atmospheric gamma rays and cosmic rays/secondary particles using the self-triggering mode of the PSA detector (Fig.~\ref{fig:flux_internal}).
In the cosmic ray/secondary particle simulation, the obtained spectrum contains different events, including direct interactions between cosmic rays/secondary particles and the PSA detector and gamma-ray events induced by interactions between the cosmic rays/secondary particles and the instrument.
We found that internal radiation was the dominant source triggering the PSAs at the balloon altitude.

\ikeda{
Cosmic rays/secondary particles activate the instrument material. 
Primarily, we suffer from the activation of the GSO scintillators, because beta and gamma rays, which are emitted in the decay process, are easily observed.
Activation of the GSO scintillator was also reported by Kokubun et al.~\cite{Kokubun1999}, and beam irradiation experiments identified various radioactive isotopes. 
To estimate the contribution of the activation background at the balloon altitude, we conducted Geant4 simulations.
The estimated count rate of the PSA scintillator was one order of magnitude less than the rate of the internal radiation in the ground experiment. 
Therefore, the contribution of the activation of the GSO scintillator was considered negligible.}

\subsection{Background event types}

\begin{figure*}
\includegraphics{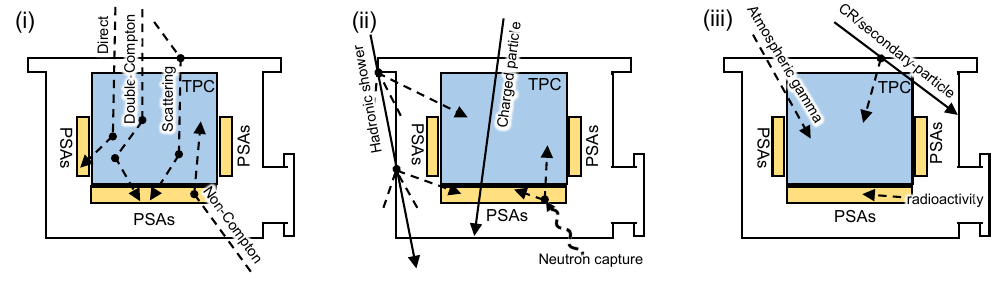}
\caption{(i) Event types of the atmospheric gamma-ray background. 
Four interaction cases corresponding to direct Compton events, double Compton events, scattering events, and non-Compton events are described.
(ii) Event types of the cosmic-ray/secondary particle background. 
Typical interaction processes such as hadronic showers, the direct incidence of charged particles, and thermal neutron capture are described.
(iii) Event types of the accidental background. The PSA and TPC are triggered by internal radioactivity and atmospheric gamma rays or cosmic rays/secondary particles, respectively.
}
\label{fig:BG_component}
\end{figure*}

The radiation discussed in Section~\ref{sec:radiation_environment} produced coincident interactions in the TPC and PSAs that passed the electronic criteria for valid gamma-ray events. 
The identified backgrounds were classified into the following types:
\begin{enumerate}
\renewcommand{\labelenumi}{(\roman{enumi})}
\item 
Atmospheric gamma-ray background: Atmospheric gamma rays were observed in several interaction cases (Fig.~\ref{fig:BG_component}(i)). 
The {\it direct-Compton event}, corresponding to the ideal Compton event, indicates that a full absorption event occurred in the PSAs after direct-Compton scattering in the TPC. 
The {\it double-Compton event} represents the case in which the Compton gamma rays deposit a part of the energy in the PSAs.
When the atmospheric gamma rays scatter off the instrumental material before entering the TPC, the events are called {\it scattering events}. 
In addition, the atmospheric gamma rays can interact with the GSO scintillator before scattering off the TPC. 
Consequently, the scattering gamma rays are absorbed or scattered in the TPC. 
Otherwise, the primary photon can generate characteristic X-rays of gadolinium that are absorbed in the TPC.
Such events do not cause Compton scattering in the TPC and are thus categorized as {\it non-Compton events}.

\item 
Cosmic-ray/secondary particle background: As the charged cosmic rays and secondary particles can deposit energy in the TPC and PSAs, they directly satisfy the electronic criteria of the ETCC. 
In addition, high-energy protons and neutrons produce multiple photons via complicated nuclear reactions such as nucleus spallation and shower initiation.
The bremsstrahlung photons are emitted by relativistic electrons and positrons, which results in the simultaneous absorption in the TPC and PSA.
In the GSO scintillator, gadolinium has a large capture cross-section for thermal neutrons owing to the contributions of  the isotopes $^{155}$Gd and $^{157}$Gd. 
The Gd($n,\gamma$) reaction between the thermal neutron and gadolinium produces four gamma rays on average~\cite{Chyzh2011}, and thus, the ETCC can possibly trigger.
A schematic of the interaction process is shown in Fig.~\ref{fig:BG_component}(ii).

\item 
Accidental background: The coincidence window of the PSA and TPC is 9.5~$\mu$s. The isolated trigger events in the PSAs and TPC that occur within the coincidence window produce an ETCC trigger signal even though the events are not physically correlated.
Thus, we consider only the case in which the PSA is triggered by internal radiation, which is the dominant trigger event in the PSA.
In contrast, atmospheric gamma rays and cosmic rays/secondary particles are evaluated based on the TPC trigger. 
The schematic is presented in Fig.~\ref{fig:BG_component}(iii).
\end{enumerate}

\section{Results}\label{sec:result}
We selected the gamma-ray events from the obtained simulation data.
The selection criteria were the same as those reported in Ref.~\cite{2022Takada}. 
However, the Compton-scattering kinematics were not adopted to assess the $\Delta\cos\alpha$ distribution.

\begin{figure}[h]
\includegraphics[width=8.6cm]{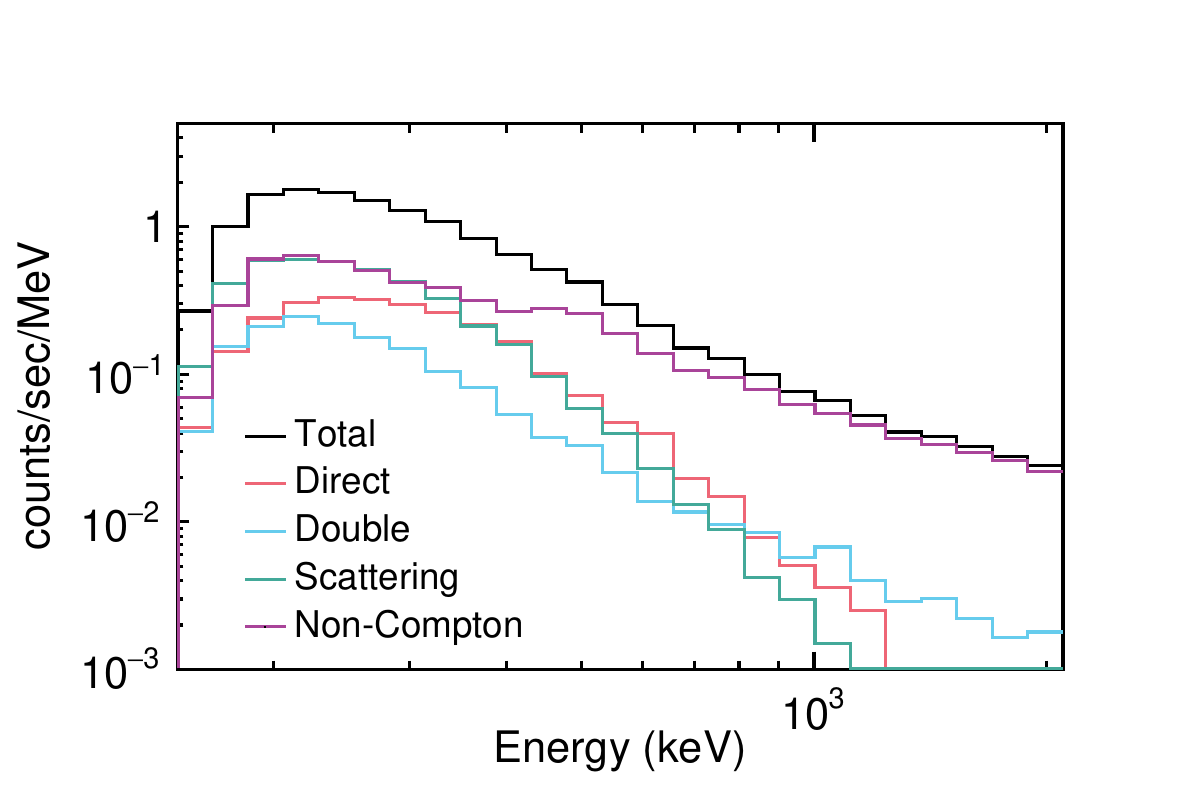}
\caption{Energy spectra results of the atmospheric gamma-ray simulations. The red, cyan, green, and purple solid lines indicate the direct-Compton, double-Compton, scattering, and non-Compton events, respectively. The black solid line indicates the total spectrum.}
\label{fig:spec_ling}
\end{figure}

The simulated spectra of the atmospheric gamma rays in the FoV are shown in Fig.~\ref{fig:spec_ling}. 
Below 400~keV, the spectrum primarily constituted the scattering and non-Compton events, whereas above 400~keV, the spectrum is dominated by non-Compton events.
Direct-Compton events were considerably reduced above 400~keV, which reflects the effective area.

\begin{figure}[h]
\includegraphics[width=8.6cm]{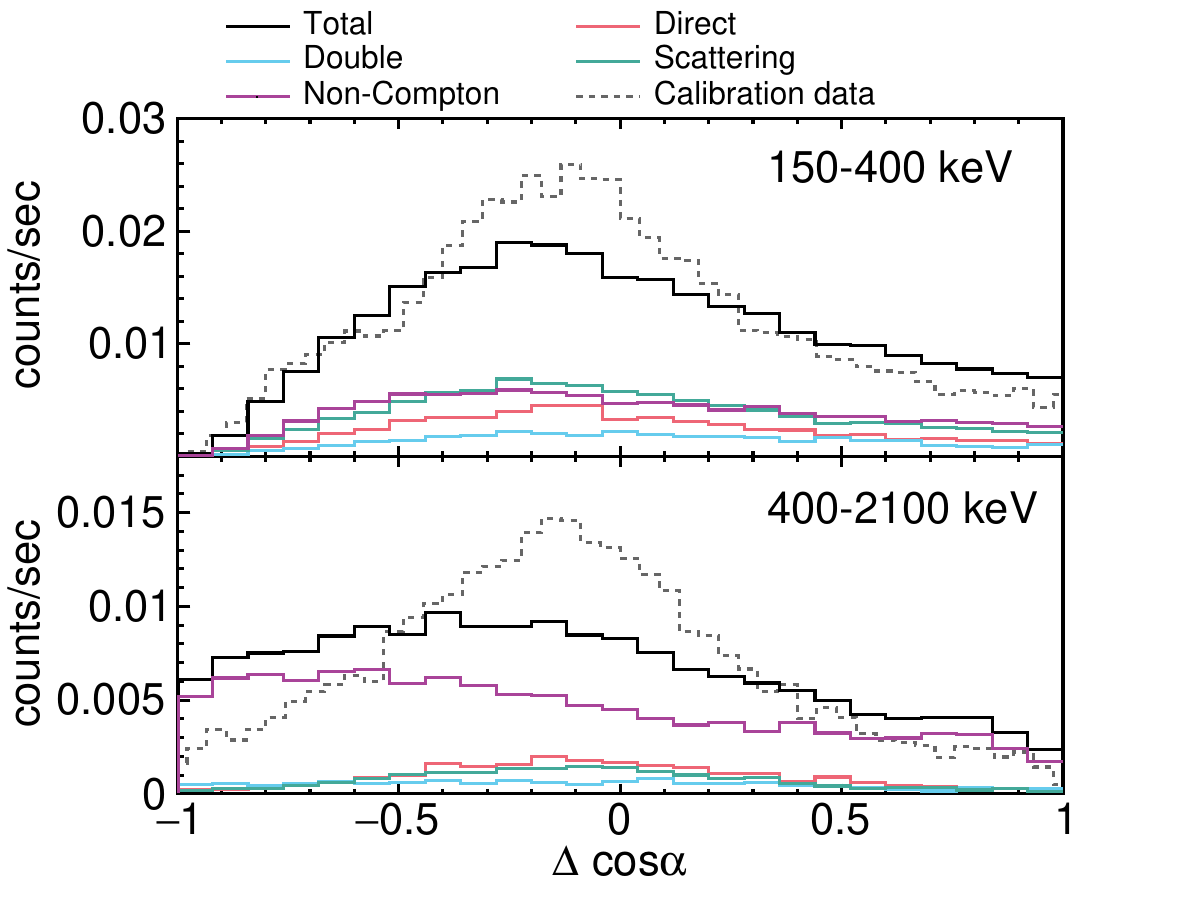}
\caption{
\ikeda{The top (bottom) panel shows the $\Delta\cos\alpha$ distribution in the range of 150--400~keV (400--2100~keV). 
The red, cyan, green, and purple solid lines indicate the direct-Compton, double-Compton, scattering, and non-Compton events, respectively. The total distribution is shown by the black line. 
The calibration data of $^{133}$Ba ($^{137}$Cs) are described in the top (bottom) panel for comparison.}
}
\label{fig:alpha_ling}
\end{figure}

The $\Delta\cos\alpha$ distribution is described in Fig.~\ref{fig:alpha_ling}. 
The distributions of the direct-Compton events and scattering events exhibited peaks near $\Delta\cos\alpha=0$, indicating ``Compton-like'' events.
Because the consideration of the high-energy recoil electron improved the accuracy of the geometrical calculation of  $\cos\alpha_{{\rm geo}}$, a sharp distribution was observed in the range of 400--2100~keV.
Conversely, the double-Compton and non-Compton events exhibited broadened distributions, suggesting that the use of $\Delta\cos\alpha$ enabled the identification of Compton-like events.

\begin{figure}[h]
\includegraphics[width=8.6cm]{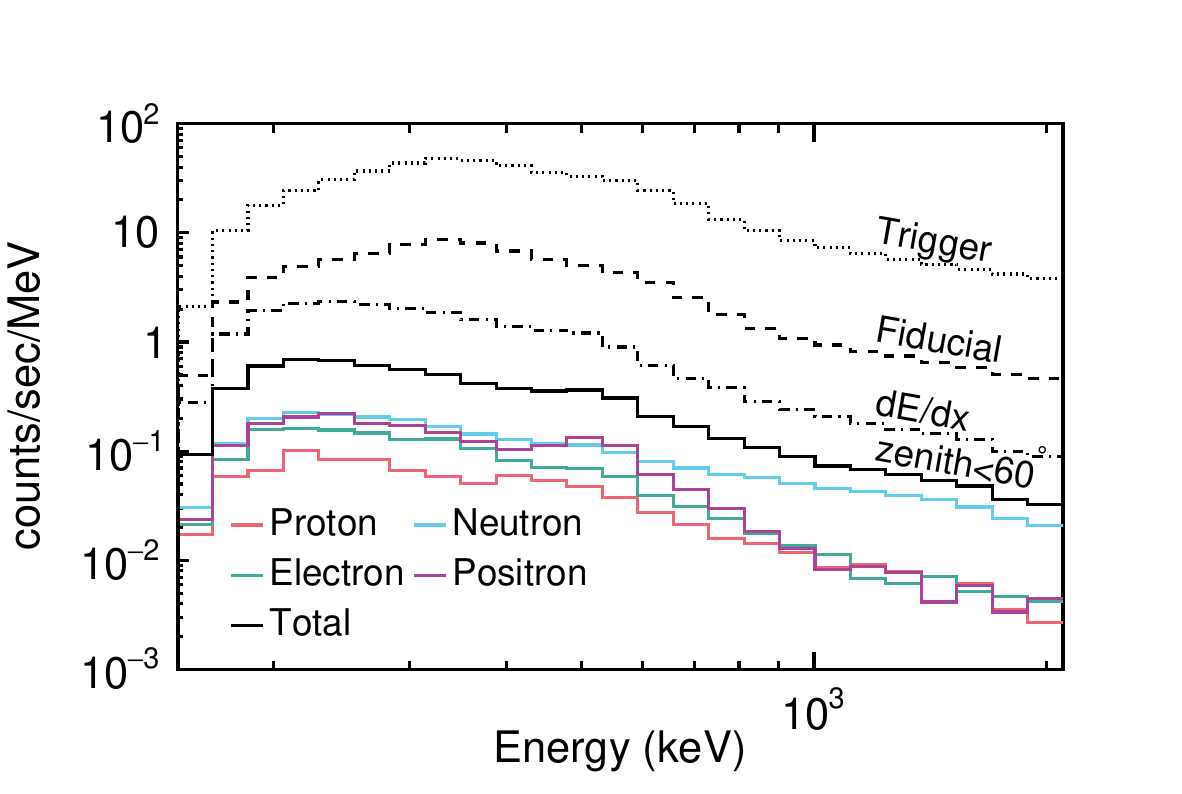}
\caption{Simulated energy spectra of the cosmic-ray/secondary-particle background. 
The black dotted, dashed, dot-dashed, and solid lines denote the triggered spectrum and the spectra after the event selection based on the fiducial, dE/dx, and zenith$<60^{\circ}$, respectively. 
The red, cyan, green, and purple spectra denote the proton, neutron, electron, and positron backgrounds, respectively.}
\label{fig:spec_CRBG}
\end{figure}

The spectra of the cosmic-ray/secondary-particle background are shown in Fig.~\ref{fig:spec_CRBG}, and the spectra are described for each selection criterion. The event selection of ``Fiducial'' and ``dE/dx'' plays a role in eliminating the charged particles passing through the TPC, thus ensuring that these selections can identify the electron track completely contained in the TPC.
We successfully reduced 96.7\% of the direct cosmic rays/secondary particles.
The neutrons, electrons, and positrons equally contributed to the remaining spectrum, whereas the neutron background dominated beyond 700~keV.
In particular, 35.3\% and 31.1\% of the neutron background were produced by the thermal neutron capture of gadolinium and the nuclear interaction cascades in the GSO scintillators, respectively. 
In the proton background, hadronic showers had the largest contribution to the interaction process, i.e., 49.1\%.
Bremsstrahlung was the dominant process in the electron and positron backgrounds, corresponding to proportions of 69.8\% and 59.7\%, respectively.

\begin{figure}[h]
\includegraphics[width=8.6cm]{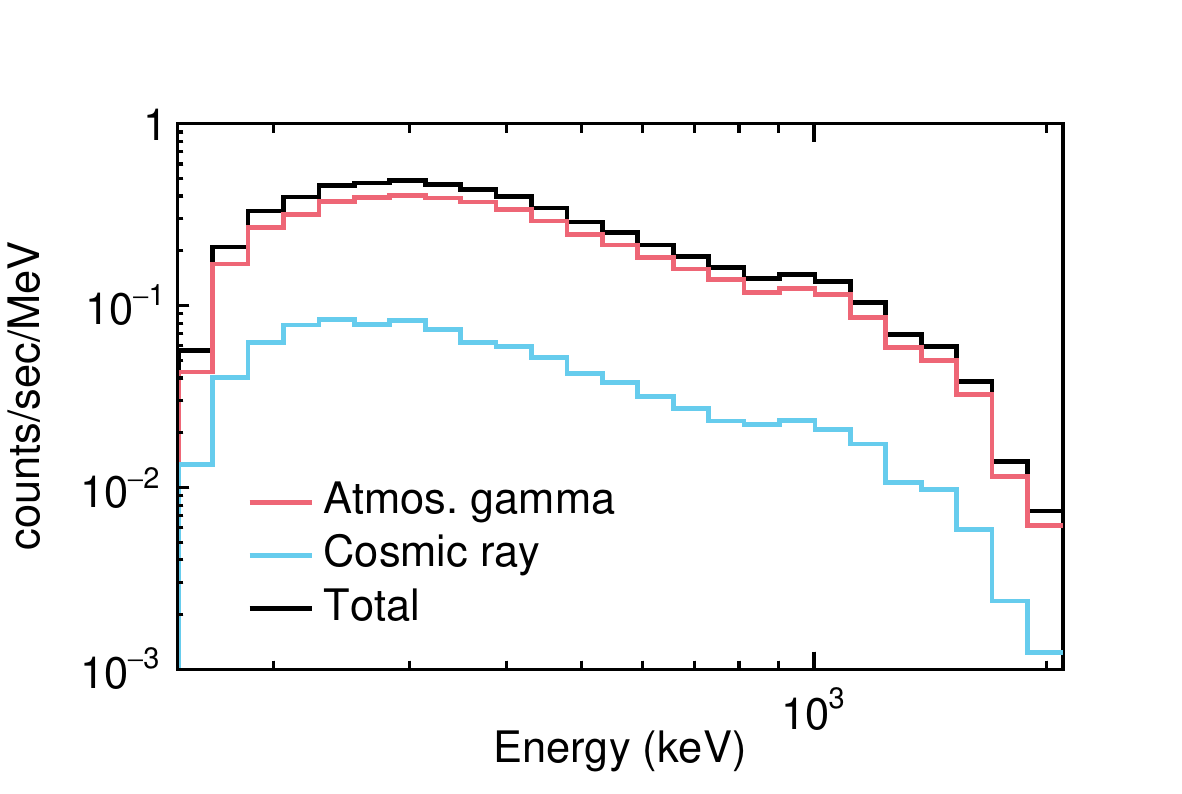}
\caption{Simulated energy spectra of the accidental background. 
The red and cyan lines indicate the contribution from the atmospheric gamma and cosmic rays, respectively, triggering the TPC. 
The black line represents the total spectrum.}
\label{fig:spec_acc}
\end{figure}

Fig.~\ref{fig:spec_acc} shows the accidental background spectrum. 
We described the contribution of the interactions between the atmospheric gamma rays and cosmic rays, including the secondary particles and induced gamma rays in the TPC. 
The accidental events caused by the atmospheric gamma rays dominated the background spectrum.
The spectrum characteristics observed at approximately 1~MeV were caused by the internal background of the alpha particles in the GSO crystals.

\begin{figure*}
\includegraphics{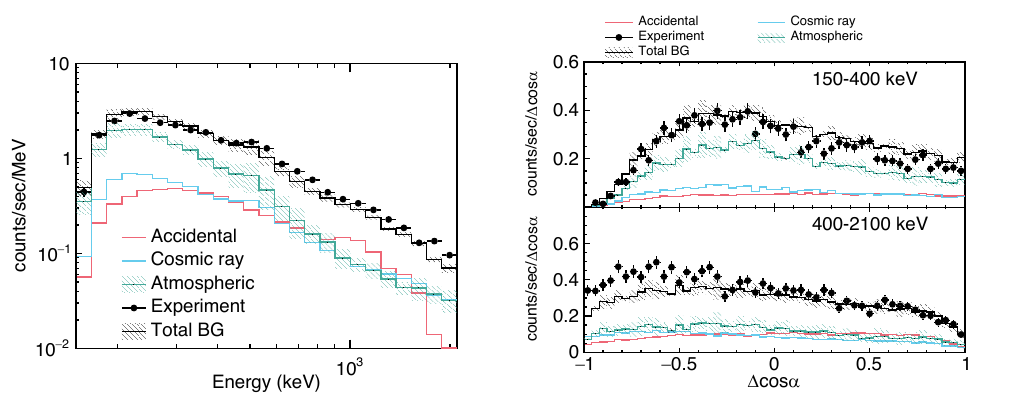}
\caption{The left and right figures show the energy spectra and $\Delta\cos\alpha$ distributions of the background simulations and the experimental data. 
The points with error bars denote the experimental spectrum of SMILE-2+. 
The red, cyan, green, and black spectra show the simulation results of the accidental, cosmic-ray, atmospheric gamma-ray, and total background, respectively.
}
\label{fig:Spectra_alpha}
\end{figure*}

Finally,  we compared the simulated background spectrum and the experimental spectrum of the steady background data. 
Fig.~\ref{fig:Spectra_alpha} shows the total background spectrum and the $\Delta\cos\alpha$ distribution obtained using the experimental data. 
Herein, we estimated the systematic uncertainty of the atmospheric gamma rays using the PARMA model, which included high-energy gamma rays above 10~MeV.
Below 400~keV, the simulation results were consistent with the experimental results within the systematic uncertainty.
However, slight discrepancies of 17\% were observed above 400~keV. 
The $\Delta\cos\alpha$ distribution in the 150--400~keV range was consistent with the experimental results.
In addition, we confirmed that one-fourth of the distribution was contributed by Compton-like events, including direct-Compton events and scattering events involving the atmospheric gamma rays, which affected the distribution around $\Delta\cos\alpha=0$.
Meanwhile, the $\Delta\cos\alpha$ distributions obtained in the experiments performed in the range of 400--2100~keV did not exhibit any peaks, implying that the primary contributions were non-Compton events involving the atmospheric gamma rays, cosmic-rays/secondary particles, and accidental backgrounds.
Furthermore, this result implies that the unexplained component in $-1<\Delta\cos\alpha<-0.3$ is unlikely to be a Compton-like event.

Herein, we compare the results of the present study with those of a previous study~\cite{2022Takada}. 
The previously simulated energy spectrum did not reproduce the bump structure at approximately 1~MeV. 
In contrast, we could replicate the spectrum when considering the accidental background.
In the previous simulated spectra, although Ling's model considered the atmospheric and cosmic diffuse gamma rays; however, in this study, we did not consider cosmic diffuse gamma rays. 
Even if we considered cosmic diffuse gamma rays, the total spectrum was within the systematic uncertainty of the atmospheric gamma rays evaluated in this study.
Additionally, in the previous study~\cite{2022Takada}, the unexplained component was effectively removed with the event selection of $-0.5<\Delta\cos\alpha<0.5$. 


\section{Discussion}\label{sec:discussion}
We discuss the remaining background candidates, which comprise 17\% of the total background in the range of 400--2100~keV.
We found that cosmic diffuse gamma rays contributed  8\% of the experimental spectrum when Ling's model was applied.
However, the remainder of the background does not potentially contain cosmic diffuse gamma rays because the corresponding $\Delta\cos\alpha$ distribution exhibits a peak at approximately zero.

Herein, we focus on the accidental background, which was evaluated based on the simulation and experimental data (Appendix~\ref{app:acc}).
The calculated count rate in the range of 400--2100~keV was higher than the simulated rate, and the differential rate was 21~$\pm$~4\% of the total experimental data. Accordingly, the remainder of the background can be fully explained. 
Moreover, the $\Delta\cos\alpha$ distribution toward negative values could be reproduced.
Thus, the remaining background was expected to be related to noncorrelated events between the TPC and PSAs.

\begin{figure}[h]
\includegraphics[width=8.6cm]{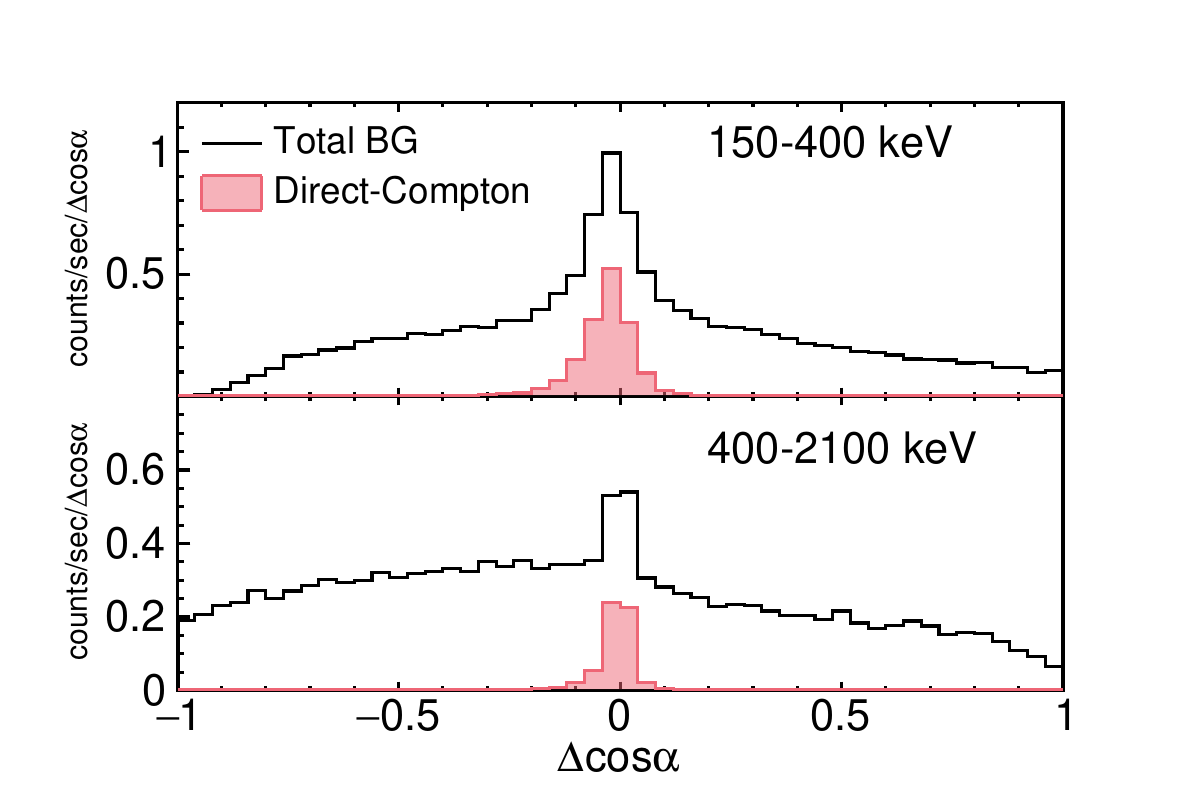}
\caption{$\Delta\cos\alpha$ distributions obtained in the background simulation assuming the true electron-recoil direction is calculated and the same energy resolution as SMILE-2+. 
The red-filled histogram indicates the direct-Compton events involving the atmospheric gamma rays. The black line shows the total background of the atmospheric gamma rays, cosmic rays, and accidental events.}
\label{fig:future_alpha}
\end{figure}

\begin{figure}[h]
\includegraphics[width=8.6cm]{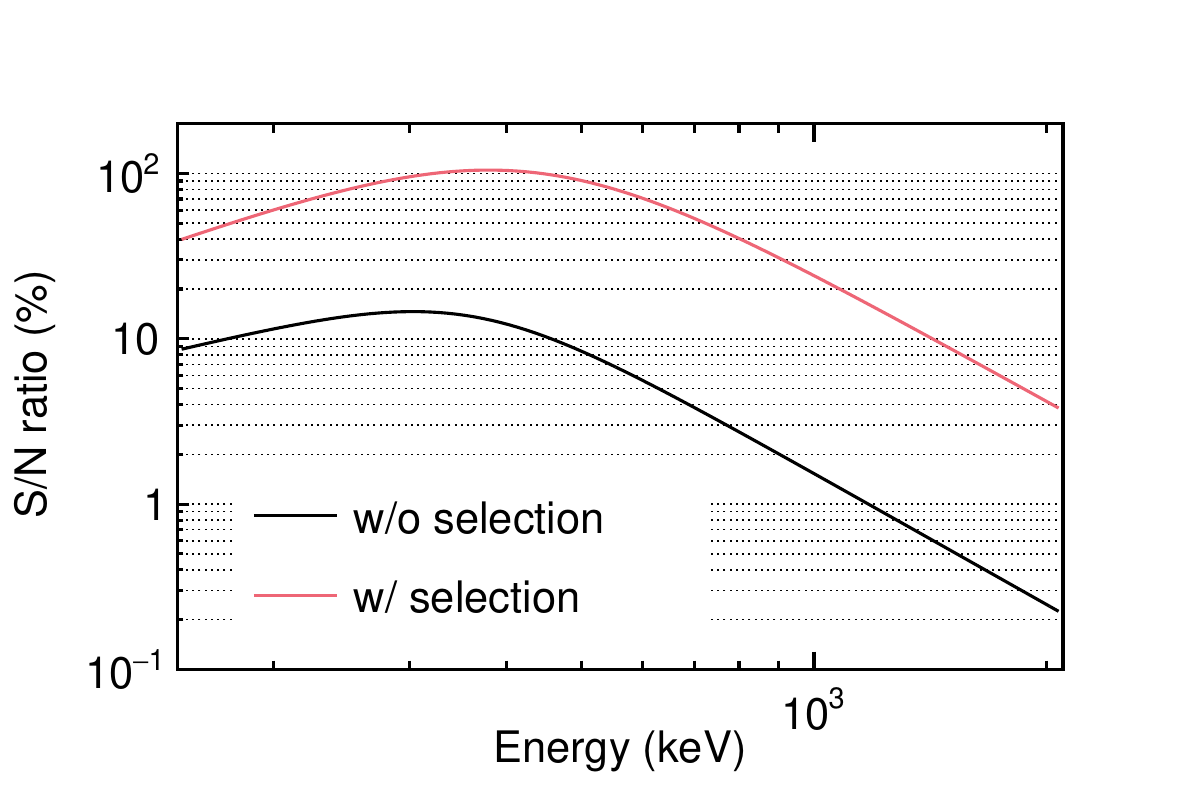}
\caption{Signal-to-noise ratio in the FoV. 
\ikeda{The black and red lines show the calculation results without and with a selection based on $\Delta\cos\alpha$.}}
\label{fig:future_ratio}
\end{figure}

Finally, we discuss the validity of the alpha kinematics test, assuming direct-Compton events with atmospheric gamma rays as the signals. 
In contrast, other events, such as non-Compton events of atmospheric gamma rays, cosmic-ray background, and accidental background, were treated as noise.
Because the determination accuracy of the electron-recoil direction in the SMILE-2+ detector is poor, the $\Delta\cos\alpha$ distribution of the direct-Compton events is broadened as shown in Fig.~\ref{fig:alpha_ling}.
However, the alpha kinematics test principally demonstrates the strong discrimination of the signal and the noise. 
Assuming that the energy resolutions of the PSA and TPC are the same as that of SMILE-2+ and that the true electron-recoil direction is obtained,
the $\Delta\cos\alpha$ distribution can be calculated, as displayed in Fig.~\ref{fig:future_alpha}. 
The appropriate selection of $\Delta\cos\alpha$ can lead to effective extraction of the signal.
The signal-to-noise (S/N) ratio in the FoV with and without the selection of $\Delta\cos\alpha$ ($|\Delta\cos\alpha| <0.05$) is shown in Fig.~\ref{fig:future_ratio}. 
The S/N ratio reached 100\% at 400~keV and was improved by an order of magnitude compared to that without the selection.
Nonetheless, the high background contamination still deteriorated the S/N ratio in the high-energy region. 
Background reduction from other perspectives, such as selecting detector materials to reduce thermal neutron capture and low-radioactive materials to reduce the accidental background, is needed.
In addition, employing an anti-counter to remove the cosmic-ray/secondary particle background associated with charged particles and implementing gamma-ray shields on the bottom of the detector to suppress non-Compton events associated with atmospheric gamma rays are effective techniques.

\section{Conclusion}
In this study, Monte Carlo simulations were conducted based on Geant4, and the background components in the ETCC on board SMILE-2+ were estimated.
Although the atmospheric gamma-ray background, the cosmic-ray/secondary particle background, and the accidental background reproduced the experimental spectrum below 400~keV, we detected the unexplained background of 17\% of the total background above 400~keV, suggesting the occurrence of unlikely Compton-like events based on the $\Delta {\rm cos}~\alpha$ distribution. 
In addition, the Compton-kinematics test considerably improved the S/N ratio by one order of magnitude. 
The ETCC detector, which can inherently discriminate Compton-like events, is expected to extend the scope of MeV gamma-ray astronomy.

\appendix
\section{Calculation of the accidental background}\label{app:acc}
\begin{figure}[!h]
\includegraphics[width=8.6cm]{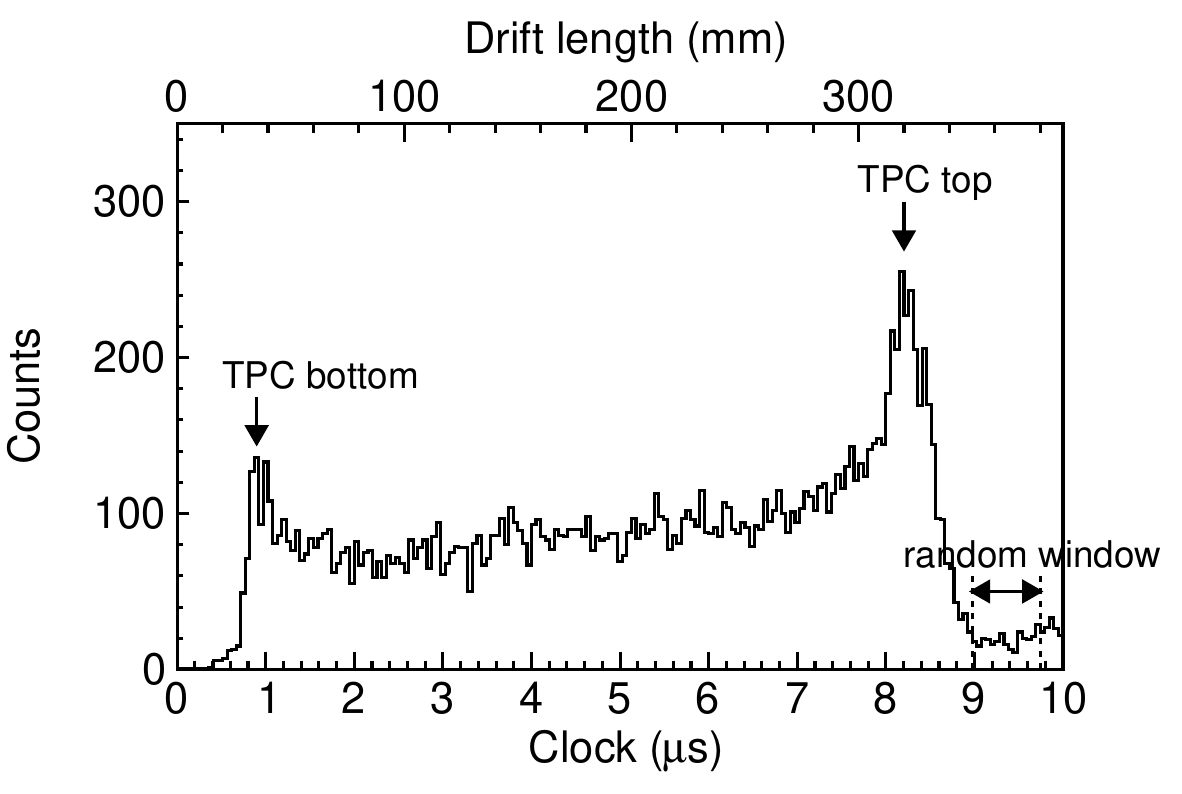}
\caption{Distribution of the maximum hit time in the TPC. The peaks observed at approximately 1~$\mu$s and 8~$\mu$s indicate the bottom and top of the TPC, respectively. 
The random window is defined as the duration between 9.0~$\mu$s and 9.7~$\mu$s.}
\label{fig:z_dist}
\end{figure}

The ETCC trigger is formed by requiring a coincidence of the signals from the PSAs and TPC. 
A gate called a coincidence window is opened after the PSA trigger, and this window should be sufficiently long to compensate for the maximum drift length. 
In SMILE-2+, we set a coincidence window of 9.5~$\mu$s considering the operated drift velocity of 3.9~cm/$\mu$s.
Fig.~\ref{fig:z_dist} shows the distribution of the acquired events with respect to the maximum value of the hit timing in TPC. 
\ikeda{Peaks appeared at approximately 1~$\mu$s and 8~$\mu$s, corresponding to the minimum drift length at the bottom of the TPC and the maximum drift length at the top of the TPC, respectively. 
The hit timing corresponding to the minimum drift length is related to the trigger timing of the PSAs.
Hence, actual Compton-scattering events occur in this time window.
The accidental events also trigger the ETCC and are distributed within the coincidence window.
The events of the hit timing above 9.0~$\mu$s are not physically correlated between the TPC and PSA. 
Thus, this window is defined as the random window in which only accidental events occur.
The accidental background could be calculated using the experimental data subtracted from the random window.}

\begin{figure}[!h]
\includegraphics[width=8.6cm]{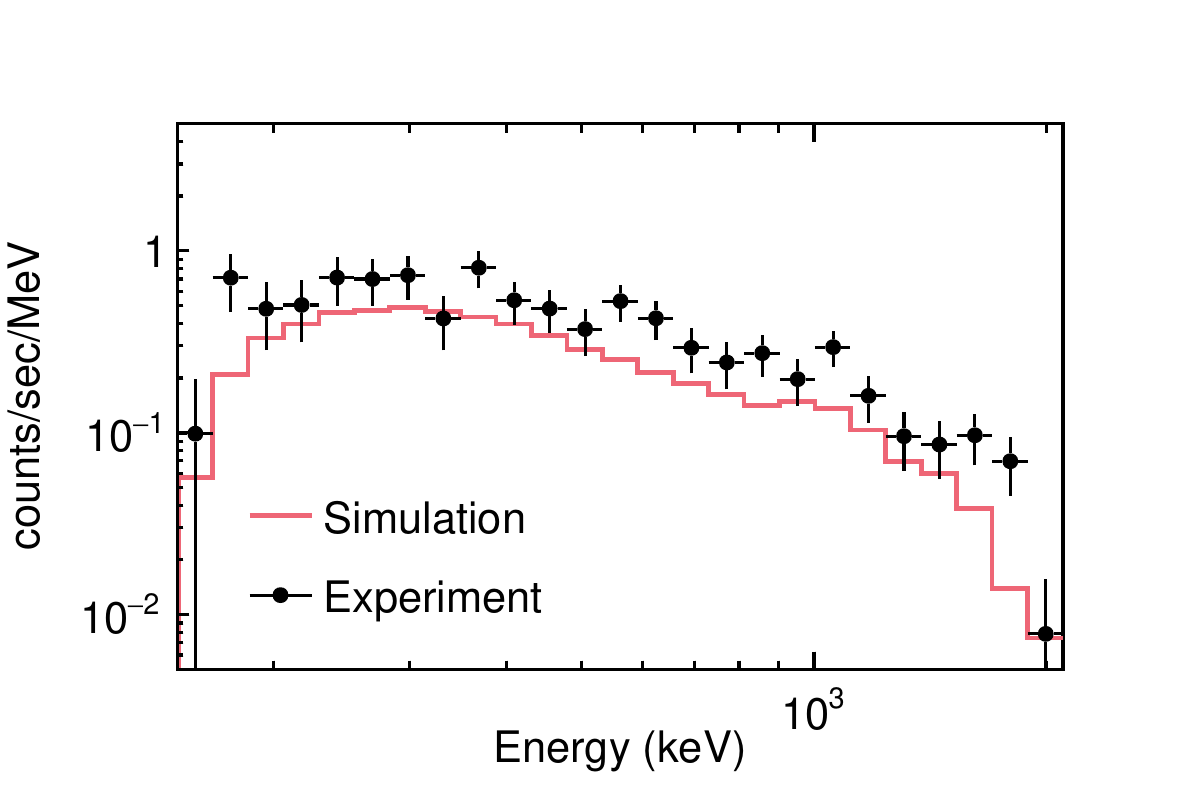}
\caption{Spectra of the accidental background. 
The red line denotes the simulation results. 
The black points and error bars indicate the experimental results obtained using the random window.}
\label{fig:comp_spec}
\end{figure}

The spectrum of the accidental background was calculated using the steady background data within the random window and is shown in Fig.~\ref{fig:comp_spec}. The spectrum was then compared with the simulation results.
The count rate above 400~keV was 322~$\pm$~23~counts/s, which was considerably different from the simulation value of 190~$\pm$~1~counts/s (only statistical error).
The count rates in the experiment and simulation below 400~keV were 167~$\pm$~17~counts/s and 111~$\pm$~1~counts/s, respectively.

\begin{figure}[h]
\includegraphics[width=8.6cm]{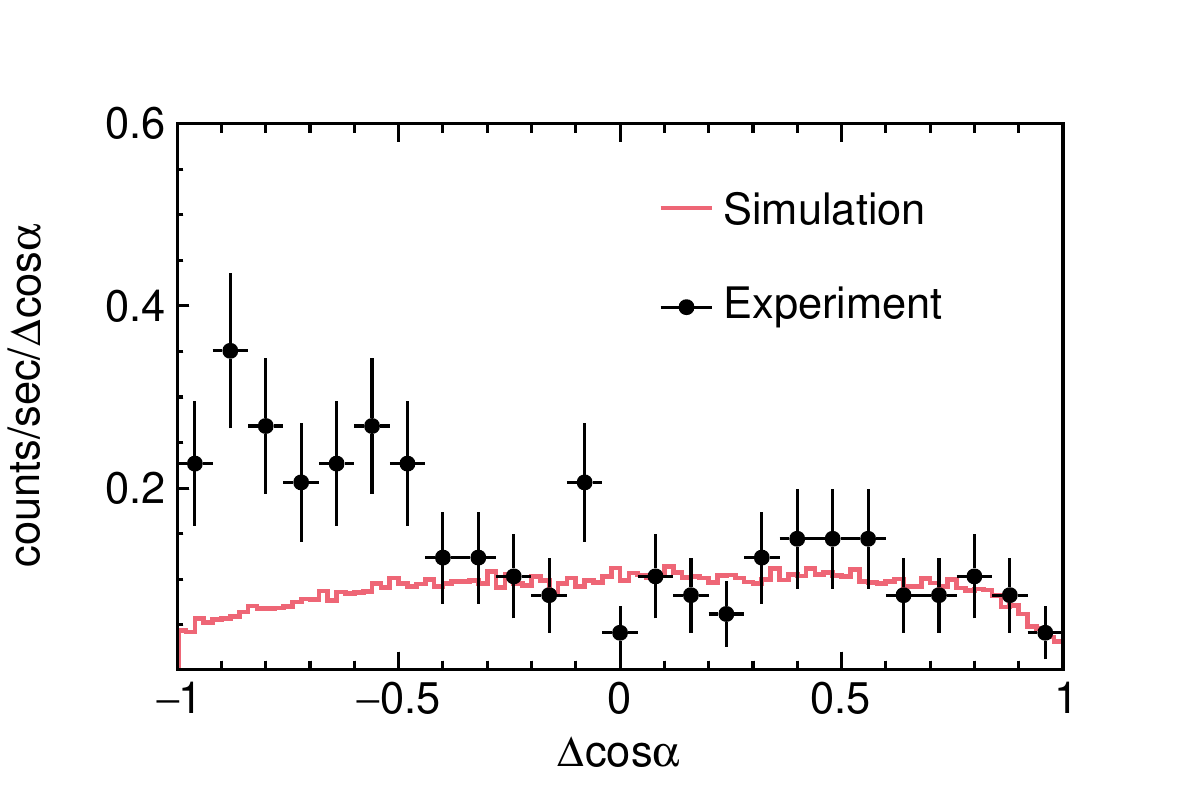}
\caption{$\Delta\cos\alpha$ distribution of the accidental background above 400~kev. 
The red line denotes the simulation results. 
The black points and error bars indicate the experimental data obtained with the random window.}
\label{fig:comp_alpha}
\end{figure}

The $\Delta\cos\alpha$ distribution of the accidental background above 400~keV obtained in the simulation is shown in Fig.~\ref{fig:comp_alpha}.
The simulation result is consistent with the experimental data for the positive values.
In contrast, the negative part is inefficient compared with the experiment.
This finding implies the existence of unexpected events, which are unlikely to be relevant to Compton-like events.

\begin{acknowledgments}
This study was supported by the Japan Society for the Promotion of Science (JSPS) KAKENHI Grant-in-Aids for Scientific Research (Grant Numbers 21224005, 20244026, 16H02185, 15K17608, 23654067, 25610042, 16K13785, 20K20428, 16J08498, 18J20107, 19J11323, and 22J00064), a Grant-in-Aid from the Global COE program “Next Generation Physics, Spun from Universality and Emergence” from the Ministry of Education, Culture, Sports, Science and Technology (MEXT) of Japan, and the joint research program of the Institute for Cosmic Ray Research (ICRR), The University of Tokyo. 
The balloon-borne experiment was conducted by researchers at Scientific Ballooning (DAIKIKYU) Research and Operation Group, ISAS, JAXA.
Some of the electronics development was supported by KEK-DTP and Open-It Consortium.
Furthermore, we would like to thank Shunsuke Kurosawa for the insightful discussion.

\end{acknowledgments}

\bibliography{apssamp}

\end{document}